\documentclass[jcap]{revtex4}

\usepackage{amsmath}
\usepackage{amsfonts}
\usepackage{amssymb}
\usepackage{color,graphicx,shortvrb,epsfig}

\begin{document}
\title{Searching for a Correlation Between Cosmic-Ray Sources Above $10^{19}$~eV and Large-Scale Structure}

\author{Tamar Kashti \& Eli Waxman}
\affiliation{Weizmann Institute of Science, Rehovot 76100, Israel}
%\today

\begin{abstract}

We study the anisotropy signature which is expected if the sources of ultra high energy, $>10^{19}$~eV, cosmic-rays (UHECRs) are extragalactic and trace the large scale distribution of luminous matter. Using the PSCz galaxy catalog as a tracer of the large scale structure (LSS) we derive the expected all sky angular distribution of the UHECR intensity. We define a statistic, that measures the correlation between the predicted and observed UHECR arrival direction distributions, and show that it is more sensitive to the expected anisotropy signature than the power spectrum and the two point correlation function. The distribution of the correlation statistic is not sensitive to the unknown redshift evolution of UHECR source density and to the unknown strength and structure of inter-galactic magnetic fields. We show, using this statistic, that recently published $>5.7\times10^{19}$~eV Auger data are inconsistent with isotropy at $\simeq98\%$ CL, and consistent with a source distribution that traces LSS, with some preference to a source distribution that is biased with respect to the galaxy distribution. The anisotropy signature should be detectable also at lower energy, $>4\times10^{19}$~eV. A few fold increase of the Auger exposure is likely to increase the significance to $>99\%$ CL, but not to $>99.9\%$ CL (unless the UHECR source density is comparable or larger than that of galaxies). In order to distinguish between different bias models, the systematic uncertainty in the absolute energy calibration of the experiments should be reduced to well below the current $\simeq25\%$.

\end{abstract}

\pacs{
98.70.Sa,    %Cosmic rays (including sources, origin, acceleration, and
            %interactions)
95.80.+p,    %Astronomical catalogs, atlases, sky surveys, databases,
            %retrieval systems, archives, etc.
98.65.Dx    %Superclusters; large-scale structure of the Universe
            %(including voids, pancakes, great wall, etc.)
}

\maketitle

\section{Introduction}

The origin of UHECRs is still  unknown: Sources have not been identified by cosmic-ray observations
and all models of particle acceleration in known astrophysical objects are challenged by the
extension of the spectrum to energies exceeding $10^{20}$~eV
\cite{Bhattacharjee:1998qc,Waxman_CR_rev}. The spectrum flattens (becomes harder)
\cite{Nagano:2000ve} and there is evidence for a composition change from heavy to light nuclei
\cite{Bird93,Abbasi05} at $\sim10^{19}$~eV. This suggests that while the cosmic-ray flux below
$\sim 10^{19}$~eV is dominated by Galactic sources of heavy nuclei, it is dominated at higher
energy by different sources of lighter nuclei. The isotropy of the UHECR arrival direction
distribution suggests that these sources are extra-Galactic.

If UHECRs are produced by a population of extra-Galactic astrophysical objects, that traces the
distribution of luminous matter, the inhomogeneous distribution of luminous matter is expected to
imprint a characteristic anisotropy signature on the UHECR arrival direction distribution. If
UHECRs are indeed light nuclei, this anisotropy is expected to be significant above
$\sim5\times10^{19}$~eV,  where the propagation distance of light cosmic-ray nuclei is limited by
interaction with the background radiation field to a distance of the order of 10's of Mpc
(\cite{Greisen:1966jv,Zatsepin:1966jv,Stecker:1999}, see \cite{Sigl01} for a detailed review). This
distance is comparable to that over which order unity variations in the density distribution of
luminous matter are observed. Identification of the expected anisotropy signature would provide
strong support to models where UHECRs are accelerated in known astrophysical objects, and would be
inconsistent with most new physics, "top-down," models, in which UHECRs are produced by the decay
of heavy relic particles or topological defects \cite{Bhattacharjee:1998qc}.

The expected anisotropy signal was analyzed in \cite{Waxman:1996hp}. It was shown there that the
signal should be detectable with high statistical significance once the number of
detected UHECR events is increased beyond that available at that time,
which was $\approx20$ events above $4\times10^{19}$~eV, by a factor $\gtrsim10$. In
this paper we revisit this subject, as the commissioning of the Auger detector may provide us
within a few years with the required exposure \cite{Yamamoto:2007xj}.
While the analysis of the current paper largely follows that of \cite{Waxman:1996hp}, it is
improved in several respects. First, for the derivation of the large-scale structure of luminous
matter we use a larger, as well as more complete and uniform, galaxy redshift survey, the PSCz catalog
\cite{Saunders:2000af} instead of the IRAS 1.2~Jy catalog \cite{Fisher:1995kd}.
The PSCz catalog contains 3 times more galaxies, including many more
galaxies at distances of $150-300$ Mpc. Second, we
use an updated cosmological model, $\{\Omega_m=0.27,\ \Omega_\Lambda=0.73,\ H_0=75\ {\rm
km/s/Mpc}\}$ instead of $\{\Omega_m=1,\ \Omega_\Lambda=0,\ H_0=100\ {\rm km/s/Mpc}\}$, and analyze
the effects of a possible redshift evolution of the UHECR source density. Third, we study the
sensitivity of the correlation statistic to systematic errors in the determination of
UHECR energies. Finally, we compare our proposed method for identifying the expected anisotropy signature to the more commonly used methods based on the power spectrum
(e.g.~\cite{Sommers:2000us,Deligny:2004dj,Anchordoqui:2003bx,Sigl}) and on the two point
correlation function (e.g.~\cite{Takeda:1999sg,DeMarco:2006a,Cuoco:2007id,Takami:2007vv} and
references therein) of the angular distribution of UHECR arrival directions.
In order to facilitate the use of the proposed statistic for the detection of anisotropy in
UHECR data, we have made available at \verb"http://www.weizmann.ac.il/"$\sim$\verb"waxman/criso" a numerical description of the intensity maps (see \S~\ref{sec:summary}).

An analysis of the expected anisotropy signal based on the PSCz catalog has recently been carried out
in \cite{Cuoco:2006}. The most important effects taken into account in the present analysis, and neglected
in ref.~\cite{Cuoco:2006}, are the dependence of the anisotropy signal on the finite number density
of sources and on the systematic errors in the determination of UHECR energies. As shown in
\S~\ref{sec:results} and discussed in \S~\ref{sec:summary}, both the finite number density of sources
and the systematic uncertainties in energy determination have important implications to the predicted
signal and its interpretation.

We assume in our analysis that the  UHERCs are protons. This assumption is motivated by two
arguments. First, the observed spectrum of $>10^{19}$~eV cosmic-rays is consistent with a
cosmological distribution of proton accelerators producing (intrinsically) a power-law spectrum of
high energy protons, $d\log n/d\log E\approx -2$, \cite{Waxman:1995dg,Bahcall:2002wi} (see also
fig.~\ref{fig:spectrum}). This intrinsic power-law spectrum is consistent with that expected in
most models of particle acceleration (\cite{Waxman_CR_rev} and references therein). Second, the
leading candidate extra-Galactic sources (GRBs and AGN, \cite{Waxman_CR_rev} and references
therein) are expected to accelerate primarily protons.

The paper is organized as follows. In sec.~\ref{sec:Method} we describe the formalism used to
analyze the anisotropy signal and present maps of the predicted angular distribution of UHECR intensity.
In sec.~\ref{sec:results} we study the sensitivity of several statistics to the predicted
anisotropy signature. In sec.~\ref{sec:Auger} we analyze Auger data, which was published during the
preparation of this manuscript \cite{AugerSc:2007},
using our correlation statistic. Sec.~\ref{sec:summary} contains a brief summary of the
results and a discussion of their implications.

The following point should be made here regarding the isotropy analysis of ref.~\cite{AugerSc:2007}. In ref.~\cite{AugerSc:2007} a correlation is found between the arrival direction
distribution of $>5.7\times10^{19}$~eV cosmic-rays (detected by the Auger experiment) and between the
angular distribution of low-luminosity AGNs included in the V-C AGN catalog \cite{V-C}. According to
the analysis of ref.~\cite{AugerSc:2007}, the probability that the detected correlation would arise
by chance for an isotropic UHECR arrival direction distribution is $\simeq1$\%. Since low luminosity AGNs
trace the distribution of luminous matter, one may argue that the detected correlation provides the first
evidence for a correlation of UHECR sources and LSS. Unfortunately, the V-C catalog is merely a
compilation of AGN data available in the literature, and is therefore incomplete both in
its sky coverage and in its luminosity coverage. It does not, therefore, provide a correct description
of the local LSS, as clearly pointed out in the introduction of ref.~\cite{V-C}: "The V-C catalog should
not be used for any statistical analysis as it is not complete in any sense, except that it is, we hope,
a complete survey of the literature". The interpretation of the correlation reported in ref.~\cite{AugerSc:2007} is thus unclear.

\section{Method}
\label{sec:Method}

We consider a model where the UHECR flux is  produced by cosmological sources of protons tracing
the large scale galaxy distribution. We assume that the sources are intrinsically identical and
that the number density of sources is drawn from a Poisson distribution with an average given by
$b[\delta]\bar{s}(z)$, where $\bar{s}(z)$ is the average comoving number density of sources at
redshift $z$ and $b$ is some bias functional of the local fractional galaxy over density,
$\delta\equiv\delta\rho/\bar\rho$. The
large scale structure galaxy density field is derived from the PSCz catalogue. We first derive in
\S~\ref{sec:map} the 2D (angular) UHECR intensity map. The effect of inter-galactic magnetic fields
on the UHECR arrival direction distribution, which is neglected in \S~\ref{sec:map}, is discussed
in detail in \S~\ref{sec:B}. We show that deflections by the inter-galactic magnetic field do not
affect significantly the UHECR intensity map on scales larger than a few degrees. In
\S~\ref{sec:MC} we describe the Monte-Carlo method used to generate realizations of the cosmic-ray
arrival direction distribution (The description given in~\ref{sec:map} and~\ref{sec:MC} is brief,
since it largely follows sections 2.1 and 2.2 of \cite{Waxman:1996hp}). In \S~\ref{sec:esimators}
we describe the statistics examined for the identification of the anisotropy signal.

\subsection{The cosmic-ray intensity map}
\label{sec:map}

We estimate the large-scale galaxy density  field by smoothing the galaxy distribution of the Point
Source Catalog redshift survey (PSCz) \cite{Saunders:2000af} with a Gaussian filter of variable
dispersion given by the the greater of $6.4$~h$^{-1}$~Mpc and the mean galaxy separation
\cite{Fisher:1995kd} (the number density of PSCz galaxies is weighted by the PSCz selection
function \cite{Saunders:2000af} before smoothing). This results in a smoothing length of
$6.4$~h$^{-1}$~Mpc out to a distance of 50~$h^{-1}$~Mpc, increasing to 10~$h^{-1}$~Mpc at a
distance of 100~$h^{-1}$~Mpc, to 26~$h^{-1}$~Mpc at a distance of 200~$h^{-1}$~Mpc and
to 50~$h^{-1}$~Mpc at a distance of 300~$h^{-1}$~Mpc. For distances greater than 300~$h^{-1}$~Mpc
we have assumed a homogeneous ($\delta\rho=0$) density field.

The PSCz catalog covers about 84\% of the sky,  where most of the unobserved area is along the
Galactic plane. This lack of coverage does not affect much the search for an extra-Galactic
anisotropy signal. In such a search it is better to avoid using the arrival distribution of UHECR
along the Galactic plane, since cosmic rays propagating through the Galactic plane may suffer
strong deflections by the Galactic magnetic field. Nevertheless, in order to present complete
intensity maps, we complete the density field in the unobserved regions following, e.g.,
\cite{Yahil:1991,Baleisis:1997wx} by adding to the catalog "galaxies" with a number drawn from a Poisson
distribution with average equal to the average density in the adjacent regions above and below the
Galactic plane.

For the bias functional we consider three models: an isotropic (I) model, in which the source
distribution is uniform, $b[\delta]=1$; an unbiased (UB) model where the source distribution traces
the galaxy distribution with $b[\delta]=1+\delta$; and a biased (B) model in which the UHECR source
distribution is biased compared to the galaxy distribution with a threshold bias,
$b[\delta]=1+\delta$ for $\delta>\delta_{\rm min}$ and $b[\delta]=0$ otherwise. We use $\delta_{\rm
min}=0$, which produces a source distribution which is more concentrated in the high density
regions of the LSS (and hence also more concentrated towards the super-galactic
plane) than IRAS galaxies are.

As discussed in detail in \cite{Waxman:1996hp}, the local mean source density, $\bar{s}(z=0)$, is
constrained by the lack of repeaters to $\bar{s}_0\equiv\bar{s}(z=0)>10^{-5}{\rm Mpc}^{-3}$. We
will therefore consider in our analysis two values, $\bar{s}_0=10^{-4}{\rm Mpc}^{-3}$ and
$\bar{s}_0=10^{-2}{\rm Mpc}^{-3}$, the latter corresponding to the number density of bright galaxies.
The redshift evolution of the source density is also unknown. We therefore consider two extreme
cases: no evolution, i.e. $\bar{s}(z)\propto (1+z)^0$, and fast evolution, $\bar{s}\propto
(1+z)^3$, corresponding to the evolution of the star-formation rate
\cite{SFR} and the AGN luminosity density \cite{Boyle:1998,QSOl} (the fastest evolution of known sources).
We include in our analysis sources out to redshift $z=2$ (the contribution of sources beyond that
redshift to the flux above $10^{19}$~eV is negligible).

\begin{figure*}[tb]
   \centerline{%\hbox{
   \includegraphics[width=0.8\textwidth]{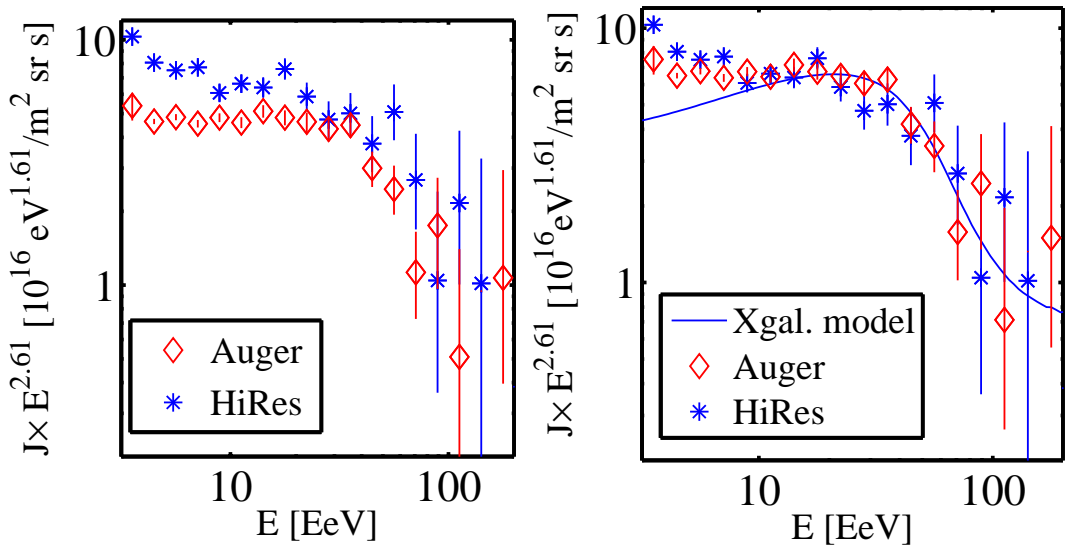}}
   \caption{The latest data on the highest energy cosmic rays. $J$
   is the differential intensity, $J=dN/dEdAdtd\Omega$.
   (a) The UHECR spectrum as published by the HiRes experiment \cite{Abbasi:2007sv} and by the Auger experiment (preliminary, \cite{Yamamoto:2007xj}). (b) The spectrum after a shift of $\Delta E/E=+23\%$ in the calibration of the absolute energy scale of the Auger experiment. The solid line is the spectrum that would be generated by a cosmological distribution of sources of protons, with intrinsic spectrum $d\log n/d\log E=-2$ and redshift evolution following the star-formation rate, $\bar{s}(z)\propto (1+z)^3$.}
    \label{fig:spectrum}
\end{figure*}
The intrinsic spectrum of protons produced by the sources is assumed to be $d\log n/d\log E=-\alpha$ with $\alpha\approx 2$, extending to $E_{\rm max}=10^{21}$~eV. As mentioned in the introduction, this is the spectrum expected in models of particle acceleration in astrophysical sources. Furthermore, it was shown in \cite{Waxman:1995dg,Bahcall:2002wi} that the UHECR spectrum produced by a cosmological distribution of sources of protons, with intrinsic spectrum $d\log n/d\log E=-2$, is consistent with the measured spectrum (as reported by the Yakutsk \cite{Glushkov:2005sc}, AGASA \cite{Takeda:2002at}, Fly's Eye \cite{Bird:1993yi} and HiRes experiments \cite{Abbasi:2007sv}). Figure~\ref{fig:spectrum} demonstrates that the preliminary spectrum of the Auger experiment is consistent with that of earlier experiments. The flux reported by the Auger experiment \cite{Yamamoto:2007xj} is significantly lower than that reported by earlier experiments, e.g. HiRes, at energies $\sim10^{19}$~eV. This discrepancy may be due to a systematic error in the absolute energy calibration of the experiments. In panel (b) of Fig.~\ref{fig:spectrum}
we have adjusted the absolute energy calibration to bring the different measured fluxes into agreement at $\sim10^{19}$~eV. This brings the measured flux into agreement at all energies $>10^{19}$~eV (There is some discrepancy remaining at lower energies, probably indicating some errors in the calculated exposures, similar to the case for earlier experiments- see detailed discussion in \cite{Bahcall:2002wi}). The relative energy shift, $|\Delta E|/E=23\%$, is well within the published systematic errors in energy scale.

\begin{figure*}[tb]
\centerline{\includegraphics[width=0.55\textwidth]{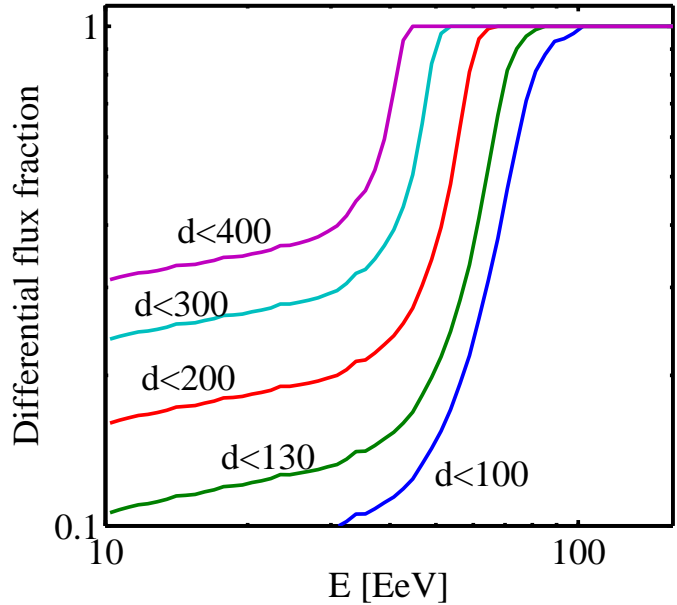}} \caption{The fraction of differential cosmic-ray flux contributed by sources  lying within a distance $d$ from Earth. A homogeneous distribution of sources of protons, with intrinsic spectrum $d\log n/d\log E=-2$ and density evolving like the star formation rate, $\propto (1+z)^3$, was assumed.}\label{fig:DiffFlux}
\end{figure*}
Figure~\ref{fig:DiffFlux} shows the contribution of sources at different distances to the differential flux of UHECRs at various energies (The contribution of sources within a given distance is lower than that presented in Fig.~1 of \cite{Waxman:1996hp} since the current calculation assumes a rapid positive redshift evolution of source density). The figure indicates that at sufficiently large energy, $\gtrsim 50$~EeV, the local, $d\lesssim100$~Mpc, inhomogeneity of matter distribution should be imprinted on the angular distribution of UHECR intensity.

\begin{figure*}[p]
   \centerline{\includegraphics[height=0.9\textheight]{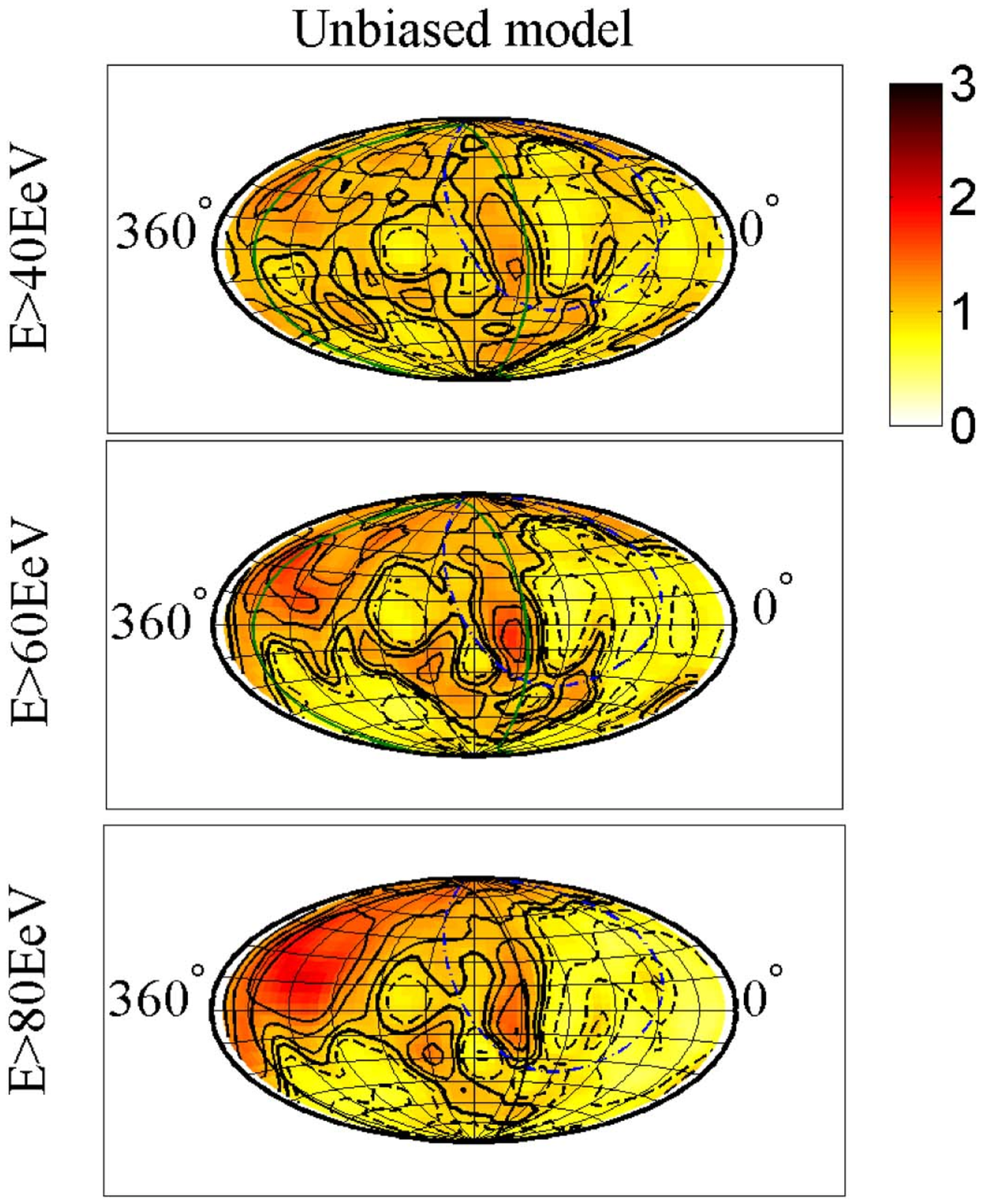}}
   \caption{All sky maps of the UHECR intensity $I(E,\hat\Omega)$, averaged over the realizations of the UHECR source distribution in the unbiased model (eq.~\ref{eq:I} with $b[\delta]=1+\delta$), for several UHECR energy thresholds $E$. The coordinates are Galactic and $I$ is normalized to its all sky average, $\bar I=\int d\Omega I(\hat\Omega)/4\pi$. Model parameters (intrinsic source spectrum $d\log n/d\log E=-2$ and redshift evolution $\propto (1+z)^3$) are similar to those of Figure~\ref{fig:DiffFlux}. The contours denote $I/\bar{I}=(0.7,0.9,1,1.1,1.3,1.5)$, with dashed lines representing under-density. The solid green line denotes the super-galactic plane. The dashed-dotted blue line marks the boundary of Auger's coverage (corresponding to a zenith angle of $60^\circ$).}
    \label{fig:CRmap1}
\end{figure*}
\begin{figure*}[p]
   \centerline{
   \includegraphics[height=0.9\textheight]{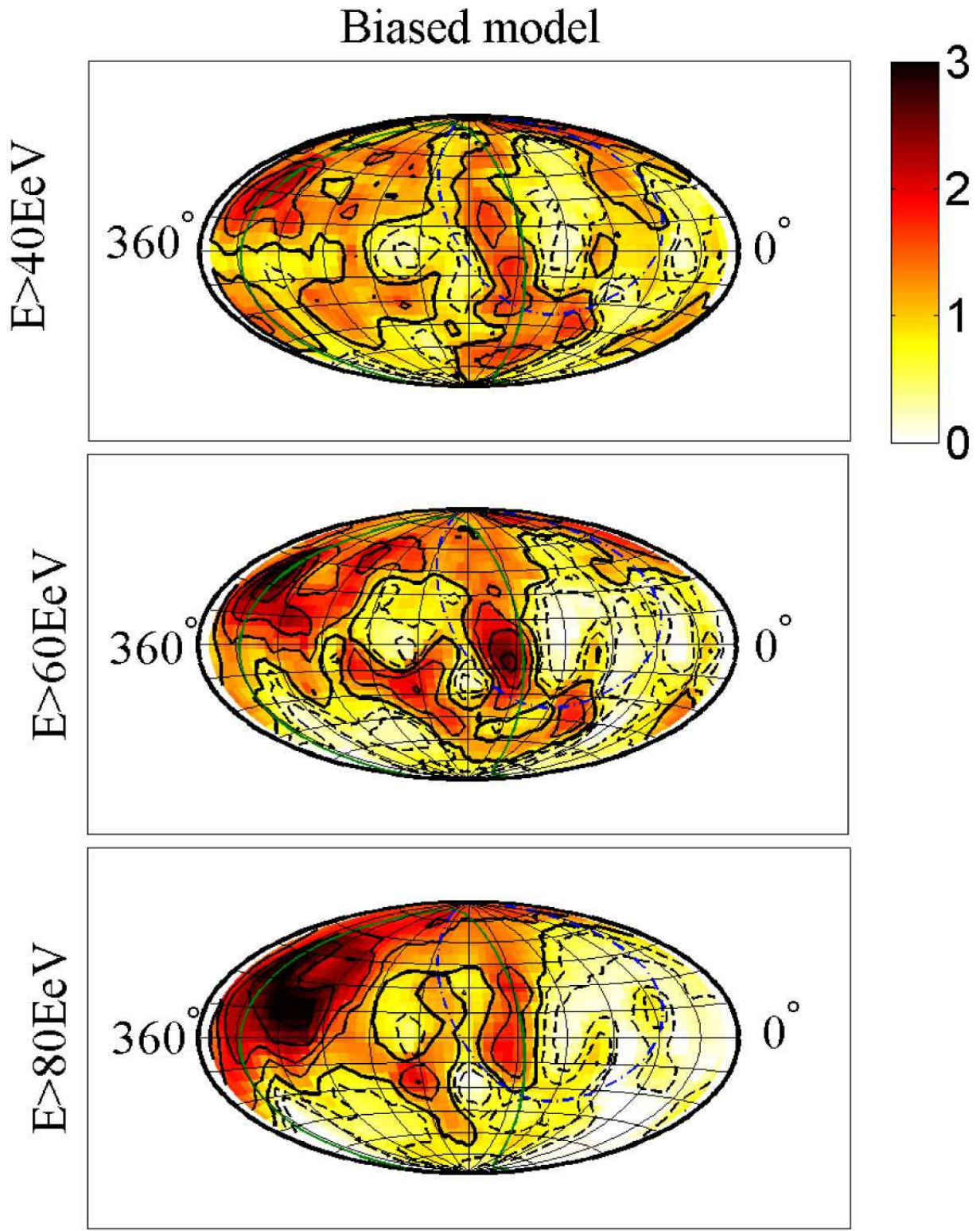}}
    \caption{Same as figure \ref{fig:CRmap1}, for the biased (B) model,
    $b[\delta]=1+\delta$ for $\delta>0$, $b=0$ otherwise.
    The contours denote $I/\bar{I}=(0.33,0.67,1,1.5,2,2.5,3).$}
    \label{fig:CRmap2}
\end{figure*}
\begin{figure*}[tb]
   \centerline{\includegraphics[width=0.6\textwidth]{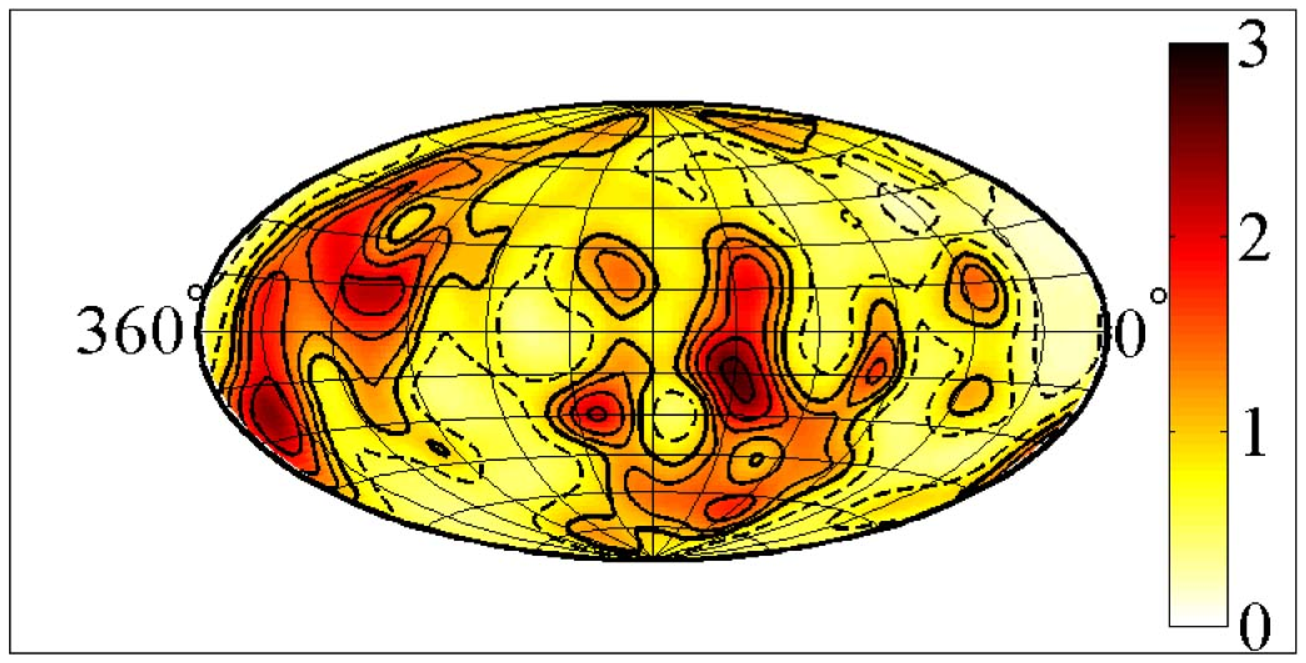}}
    \caption{The integrated galaxy density out to a distance of $75$~Mpc,
    normalized to the mean integrated density. The contours are logarithmic, ranging from 0.5 to 4
    with three contours per density doubling. Dashed curves represent under-density.}
    \label{fig:densityMap75}
\end{figure*}

Figures~\ref{fig:CRmap1} and~\ref{fig:CRmap2} present all sky maps of the UHECR intensity $I(E,\hat\Omega)$, averaged over the realizations of the UHECR source distribution in the
unbiased and biased models, for several UHECR energy thresholds $E$. $I(E,\hat\Omega)$ is given by \cite{Waxman:1996hp}
\begin{align}\label{eq:I}
    I(E,\hat \Omega)=\frac{\dot{n}_0(E)}{4\pi}\int dz c\left|\frac{dt}{dz}\right|
    b\left[\delta(z,\hat \Omega)\right]\frac{\bar s(z)}{\bar s_0}\frac{\dot{n}_0[E_0(E,z)]}{\dot{n}_0(E)}.
\end{align}
Here, ${\dot n}_0(E)$ is the average number of protons of energy larger than $E$ produced per unit
volume and time at $z=0$, and $E_0(E,z)$ is the energy with which a proton should be produced at redshift $z$ in order for it to be observed at $z=0$, following energy loss due to interaction with the microwave background radiation, with energy $E$ (We calculate $E_0(E,z)$ using the continuous energy loss approximation, following \cite{Waxman:1995dg}). The angular structure of the UHECR intensity map reflects the local large scale structure of the galaxy distribution, as can be seen by comparing figures~\ref{fig:CRmap1} and~\ref{fig:CRmap2} to figure~\ref{fig:densityMap75}, which presents the integrated galaxy density out to a distance of 75~Mpc. One may clearly identify the "Great Attractor", composed of the Hydra-Centaurus ($300^\circ<l<360^\circ$, $0^\circ<b<45^\circ$) and Pavo-Indus ($320^\circ<l<360^\circ$, $-45^\circ<b<0^\circ$) super clusters, the Perseus-Pisces super cluster ($130^\circ<l<160^\circ$, $-30^\circ<b<30^\circ$), the NGC1600 group ($l\sim200^\circ$, $b\sim-30^\circ$), and the local voids along the Galactic plane (at $0^\circ<l<120^\circ$).

\subsection{Magnetic fields}
\label{sec:B}

UHECR may suffer significant deflections as they  cross dense large scale structures, like galaxy clusters and large scale galaxy filaments, in which the energy density of the plasma is large enough to support strong magnetic fields. Let us consider first deflections by massive clusters of galaxies. As pointed out in \cite{WL00}, observed magnetic fields are consistent with the assumption that the magnetic field energy density amounts to $\sim1\%$ of the plasma thermal energy density (i.e. 1\% of equipartition). For a typical temperature of a few keV and density of $\sim10^{-4}{\rm cm}^{-3}$ over a scale of 1~Mpc, this gives $B\simeq0.1\,\mu$G, consistent with observations  \cite{Kim:1989,Rephaeli:1999,Hwang:1997}. The
deflection of an UHECR passing through a cluster is approximately given by
\cite{MiraldaEscude:1996kf}
\begin{align}\label{eq:DeflectionCluster}
\theta\approx\left(\frac{2}{9}\right)^{1/2} \frac{(D\lambda)^{1/2}}{E/eB} = 2.5^\circ
\left(\frac{L}{1~\mbox{Mpc}}\frac{\lambda}{10~\mbox{kpc}}\right)^{1/2}
\left(\frac{\epsilon_B}{0.01}\right)^{1/2}\left(\frac{E}{10^{20}{\rm eV}}\right)^{-1},
\end{align}
where $\epsilon_B$ is the ratio of magnetic field to thermal plasma energy density ($B\sim\sqrt{8\pi\epsilon_B nT}$), $L$ is the propagation distance through the magnetized plasma,  and $\lambda$ is the field correlation length, which is typically inferred from Faraday rotation measures to be of order 10~kpc \cite{Kim:1990,Feretti:2003}. Thus, a deflection of a few degrees is possible
if the UHECR passes through a massive cluster. However, massive clusters are very rare objects, which contain a negligible fraction of the mass in the universe and cover only a negligible fraction of the sky.

Passing through a large scale filament is more likely, but produces much smaller deflections: For a typical density of $\sim10^{-6}{\rm cm}^{-3}$ over a scale of 1~Mpc and temperature of $\sim0.1$~keV \cite{Klypin:2001ai}, the magnetic field is expected to be (assuming 1\% of equipartition) $B\sim 5$~nG leading to a deflection
\begin{align}\label{eq:DeflectionFilament}
\theta\approx 0.1^\circ\left(\frac{L}{1~\mbox{Mpc}}\frac{\lambda}{10~\mbox{kpc}}\right)^{1/2}
\left(\frac{\epsilon_B}{0.01}\right)^{1/2}\left(\frac{E}{10^{20}{\rm eV}}\right)^{-1}.
\end{align}
Assuming that a fraction $f\sim0.1$ of the volume is filled with filaments of diameter $L\sim1$~Mpc, the number of filaments crossed by an UHECR propagating a distance $d$ is $\simeq fd/L$, and the total deflection is thus
\begin{align}\label{eq:TotalDeflectionFilament}
\theta\approx 0.3^\circ\frac{L}{1~\mbox{Mpc}}\left(\frac{f}{0.1}\frac{d}{100~\mbox{Mpc}}
\frac{\lambda}{10~\mbox{kpc}}\right)^{1/2}
\left(\frac{\epsilon_B}{0.01}\right)^{1/2}\left(\frac{E}{10^{20}{\rm eV}}\right)^{-1}.
\end{align}

The results derived above are consistent with the recent semi-analytic analysis of \cite{Lemoine08}, who obtain $\theta\approx 3^\circ$ for $E=10^{20}$~eV and $d=100$~Mpc, assuming $\epsilon_B=0.1$ and $\lambda=100$~kpc. The values adopted in \cite{Lemoine08} for $\epsilon_B$ and $\lambda$, and hence the deflection $\theta$ derived in \cite{Lemoine08}, should be considered as upper limits: As noted above, $\epsilon_B=0.1$ is a value larger than typically inferred, and building an equipartition field with $\lambda=100$~kpc coherence scale is marginal since the turn over time of $\lambda=100$~kpc eddies in the expected LSS flow is comparable to the Hubble time.
Our results are also consistent with, and provide a simple explanation of, the results of detailed numerical calculations. Dolag et al.~\cite{Dolag:2004kp} have used numerical simulations of large scale structure formation, constrained to reproduce the large scale structure determined by the IRAS catalog out to $\sim100$~Mpc, in order to study the expected deflection of UHECR by inter-galactic magnetic
fields. Assuming that the magnetic field energy density constitutes a significant fraction of the
(turbulent and thermal) energy density of the inter-galactic plasma, they conclude that the
deflection of $\ge40$~EeV UHECR does not exceed 2~degrees over 99\% of the sky.

The deflection of $4\times10^{19}$~eV protons over $d\sim100$~Mpc propagation distance is therefore expected to be smaller than or comparable to a few degrees. Note, that although the deflection may be somewhat larger for longer propagation distance, the largest propagation distance of $>4\times10^{19}$~eV protons is $\sim350$~Mpc (see fig.~\ref{fig:DiffFlux}), the random deflections of particles originating from sources at distances $>100$~Mpc are not expected to affect significantly the intensity distribution:
The anisotropy of the angular UHECR intensity distribution is dominated by structures at distances $d\lesssim100$~Mpc, while sources at larger distances produce a roughly isotropic "background". We therefore expect deflections not to modify the intensity distribution of $>4\times10^{19}$~eV on scales larger than a few degrees.

It should be pointed out here that large distortions of the UHECR intensity map could have been generated by inter-galactic magnetic fields had the Milky Way galaxy been
embedded in a high density large scale filament or cluster, which may support a large magnetic
field. However, both observations and constrained simulations of our local super-cluster, e.g.
\cite{Klypin:2001ai}, show that the Galaxy does not lie within or near such a structure. The
discrepancy between the results of \cite{Dolag:2004kp} and those of \cite{Sigl:2003ay}, who
predict much larger deflections based on unconstrained large scale structure simulations, may
largely be due to the presence of highly magnetized large scale structures near, or around, the
Galaxy, which are inconsistent with the observed nearby galaxy distribution (see \cite{Dolag:2004kp} for detailed discussion).

\subsection{Monte-Carlo realizations}
\label{sec:MC}

Following \cite{Waxman:1996hp} we denote by $S_i(E,\hat \Omega)$ the number of sources per unit solid angle in the direction $\hat\Omega$, that produce $i$ events in the detector above energy $E$. It was shown in \cite{Waxman:1996hp} that $S_i$ and $S_j$ are statistically independent for $j\neq i$, and that $S_i$ is Poisson distributed with an average
\begin{align}\label{eq:Si}
    \bar S_i(E,\hat \Omega)=\int dz c\left|\frac{dt}{dz}\right|\frac{d_L(z)^2}{1+z}\bar s(z)
    b\left[\delta(z,\hat \Omega)\right] P_i(E,z).
\end{align}
Here, $d_L$ is the luminosity distance and  $P_i(E,z)$ is the probability that a source at
redshift $z$ produces $i$ events above energy $E$. $P_i$ is given by a Poisson distribution,
\begin{equation}\label{eq:Pi}
   P_i(E,z)=\frac{\bar N(E,z)^i}{i!}e^{-\bar N(E,z)},
\end{equation}
where $\bar N(E,z)$ is the average number of events above energy $E$ produced by a source at redshift $z$. $\bar N(E,z)$ is given by the product of the effective detector area, $A$, the observation time, $T$, and the flux $F(E,z)$ of particles above energy $E$ produced by a source at redshift $z$, $\bar  N(E,z)=F(E,z)AT$. $F(E,z)$ is related to the rate $\dot N(E)$ with which particles of energy larger than $E$ are produced by a single source by \cite{Waxman:1996hp}
\begin{align}
  F(E,z)=(1+z)\frac{\dot N[E_0(E,z)]}{4\pi d_L(z)^2}.
\end{align}
Here, $E_0(E,z)$ is the energy with which  a proton should be produced at redshift $z$ in order
for it to be observed at $z=0$, following energy loss due to interaction with the microwave background radiation, with energy $E$. Expressing $\dot N(E)$ as a ratio between ${\dot n}_0(E)$, the average number of protons of energy larger than $E$ produced per unit volume and time at $z=0$, and ${\bar s}_0$, we finally have \cite{Waxman:1996hp}
\begin{align}\label{eq:Nbar}
    \bar N(E,z)=\frac{\dot n_0\left[E_0(E,Z)\right]}{{\bar s}_0}\frac{(1+z)A T}{4\pi d_L(z)^2}.
\end{align}
Note, that the effective area and observing  time of a particular experiment, $AT$, may in general
depend on $\hat\Omega$, in which case $\bar N$ and $P_i$ become $\hat\Omega$ dependent as well,
$\bar N(E,z,\hat\Omega)=F(E,z)AT(\hat\Omega)$.

Using eqs.~(\ref{eq:Si}), (\ref{eq:Pi}) and~(\ref{eq:Nbar}) we generate a realization of the cosmic-ray arrival direction distribution by drawing the number of sources producing $i$ events above energy $E$ in an angular region $\rm{d}\Omega$ around direction $\hat\Omega$ from a Possion distribution with average $\bar S_i(E,\hat\Omega)\rm{d}\Omega$. The production rate $\dot n_0$ is normalized based on the observed
UHECR flux. The normalization is uncertain due to the uncertainty in the absolute energy calibration of the UHECR experiments (\cite{Bahcall:2002wi}, see also fig.~\ref{fig:spectrum}). Adopting, for example, the absolute energy scale of the HiRes experiment, the average number of events detected above 40~EeV is 300 per year for the Auger exposure of $7000$~km$^2$~sr, and about 50 events per year for the planned Telescope Array exposure of $1200$~km$^2$~sr \cite{TA:2009}. These numbers are reduced by a factor of approximately 2 if the preliminary absolute energy scale of the Auger experiment is chosen (see fig.~\ref{fig:spectrum}), and increased by a factor of about 1.5 if the AGASA absolute energy scale is adopted (see fig.~1 of \cite{Bahcall:2002wi}).

We take into account the statistical uncertainty in the energy determination of the experiment by modifying $\bar N(E,z)$: $\bar N(E,z)$ is replaced with $\int dE' [d\bar N(E',z)/dE'] g(E',E)$ where $g(E',E)$ is the probability that an event of energy $E'$ will be measured to have energy exceeding $E$. This procedure correctly reproduces the average observed number of events, $\bar N(E,z)$, but leads to some overestimate of the variance of the observed number of events. For a relative statistical energy uncertainty $\Delta E/E$, we assume that the measured energy $E$ follows a Gaussian distribution around the true energy $E'$, with a standard deviation of $\Delta E(E')$. As we show in \S~\ref{sec:results}, the statistical uncertainty in energy determination does not affect the results significantly.

\subsection{Measures of anisotropy}
\label{sec:esimators}

We consider several methods for the detection of the anisotropy of the cosmic-ray arrival distribution, using statistics based on the two point correlation function, on the angular power spectrum, and on a correlation between the observed and predicted arrival direction distributions. The various statistics are described below. In \S~\ref{sec:results} we compare the sensitivity of these statistics to the expected anisotropy signal by deriving, using the Monte Carlo simulations, their expected distributions for the different models (isotropic, unbiased, and biased UHECR source distribution).

The (cumulative) two-point correlation function is defined as the number of pairs of events that have a distance smaller then an angular distance $D$ (e.g.~\cite{Kachelriess:2005uf}),
\begin{align}\label{eq:W}
W(D)=\sum_i^N\sum_{j<i}\Theta(D-D_{ij})
\end{align}
where $\Theta$ is the step function, $N$  is the number of CR events, and $D_{ij}$ is the angular
distance between CR events $i$ and $j$. We use  several statistics based on the two-point correlation function, all of the form
\begin{align} \label{eq:X_W}
  X_W(\{D_i\})\equiv \Sigma_{i}
  \frac{\left[W(D_{i},D_{i+1})-W_{iso}(D_{i},D_{i+1})\right]^2}{\sigma^2(D_{i},D_{i+1})},
\end{align}
where $\{D_i\}$ is a set of angular distances,  $D_{i}<D_{i+1}$, $W(D_{i},D_{i+1})\equiv W(D_{i+1})-W(D_{i})$, $W_{iso}(D_{i},D_{i+1})\equiv W_{iso}(D_{i+1})-W_{iso}(D_{i})$, $W_{iso}(D)$ is the average value of the correlation function obtained in the isotropic model, and $\sigma^2(D_{i},D_{i+1})$ is the variance of $W_{iso}(D_{i},D_{i+1})$ obtained in the isotropic model.

The 2D angular UHECR intensity map may be decomposed into spherical harmonics, $I(\hat\Omega)=\sum_{\ell
m}a_{\ell m}Y_{\ell m}(\hat\Omega)$. The observed UHECR arrival direction distribution may be used
for estimating the coefficients $a_{\ell m}$ by \cite{Tegmark:1995hi}
\begin{align}\label{eq:alm}
  a_{\ell m} = \frac {1} {\mathcal{N}} \sum _{i=1}^N \frac{1}{w_i}Y_{\ell m} (\hat\Omega^i).
\end{align}
Here $N$ is the number of discrete arrival directions $\{\hat\Omega^i\}$, $\mathcal{N}=\sum_{i=1}^N 1/w_i$, and $w_i$ is the relative experimental exposure at arrival direction $\hat\Omega_i$ (e.g. \cite{Sommers:2000us}). The angular power spectrum coefficients are given by
\begin{align}\label{eq:cl}
  C_\ell = \frac{1}{2\ell+1}\sum_{m=-\ell}^{m=\ell} a_{\ell m}^2.
\end{align}
We use several statistics based on the two-point correlation function, all of the form
\begin{align}
 X_Y(\{\ell\})=\Sigma_{\ell}\frac{(C_\ell-C_{iso,\ell})^2}{\sigma_\ell^2},
\end{align}
where $\{\ell\}$ is a set of $\ell$ values, $C_{iso,\ell}$ is the average value
of $C_\ell$ obtained in the isotropic model, and $\sigma^2_\ell$ is the variance of $C_\ell$
obtained in the isotropic model.

Finally, we consider a statistic characterizing the correlation between the predicted and observed
arrival distributions,
\begin{align}\label{eq:X_C}
   X_{C,M}=\sum_{\{i\}}\frac{(N_i-N_{i,iso})(N_{i,M}-N_{i,iso})}{N_{i,iso}}.
\end{align}
Here $\{i\}$ is a set of angular bins, $N_i$ is the number of events detected in bin $i$, $N_{i,iso}$ is the average number of events expected to be detected in an isotropic model, and $N_{i,M}$ is the average number of events expected to be detected in the $M$ (e.g. unbiased, biased) model. Throughout the paper we calculate $X_C$ using $6^\circ\times6^\circ$ angular bins, in order to avoid sensitivity to the possible distortion of the arrival direction distribution by inter-galactic magnetic field deflections (see \S~\ref{sec:B}).

\section{Results}
\label{sec:results}

\begin{figure*}[tb]
   \centerline{\includegraphics[width=0.9\textwidth]{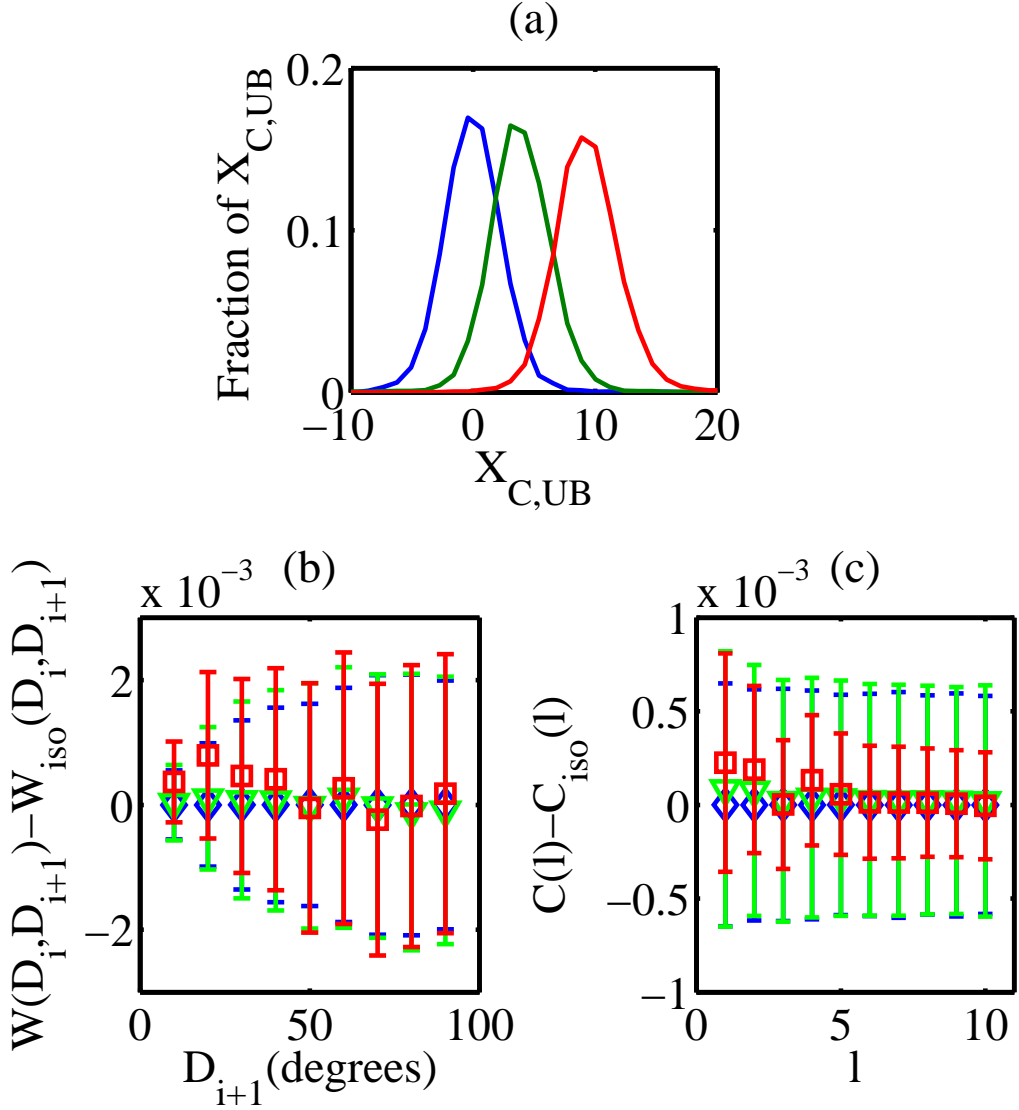}}
\caption{Properties of the distributions of $X_C$, $W(D_{i},D_{i+1})$ and $C_\ell$ obtained from 10000 Monte-Carlo realizations of 300 events above $4\times10^{19}$~eV for the isotropic model (blue diamonds), unbiased model (green triangles, $b[\delta]=1+\delta$) and biased model (red squares, $b[\delta]=1+\delta$ for $\delta>0$, $b[\delta]=0$ otherwise):
    (a) Probability distribution of $X_{C,UB}$;
    (b) Average and 1~$\sigma$ range of $W(D_{i},D_{i+1})-W_{iso}(D_{i},D_{i+1})$ as a function of $D_{i+1}$;
    (c) Average and 1~$\sigma$ range of $C_\ell-C_{iso,\ell}$ as a function of $\ell$.
We used here $\bar{s}_0=10^{-4}{\rm Mpc}^{-3}$, $\bar s\propto(1+z)^3$, $d\log n/d\log E=-2$, $\Delta E/E=0.2$, and a uniform sky exposure.}
    \label{fig:comp}
\end{figure*}

Let us first compare the sensitivity of the various statistics to the expected
anisotropy signal. For this comparison we adopt a local ($z=0$) source density
$\bar{s}_0=10^{-4}{\rm Mpc}^{-3}$, a source density evolution following the star-formation rate,
$\bar s\propto(1+z)^3$, and an intrinsic source spectrum $d\log n/d\log E=-2$. We consider the
angular arrival direction distribution of 100 cosmic-rays above $40$~EeV, detected by an experiment
with a statistical uncertainty of (single) event energy determination of $\Delta E/E=20\%$. We
consider a (hypothetical) detector with uniform exposure over the sky, and a detector with an
angular exposure pattern corresponding to that of the Auger experiment (as given
in  \cite{Sommers:2000us}) or of the Telescope Array experiment \cite{TA:2009}. For the current comparison
we do not exclude the Galactic plane strip.

Fig.~\ref{fig:comp} presents properties of the distributions of $X_{C,UB}$ (eq.~\ref{eq:X_C}),
$W(D_i,D_{i+1})$ (eq.~\ref{eq:W}) and $C_l$ (eqs.~\ref{eq:alm} \&~\ref{eq:cl}), obtained from 10000 Monte-Carlo realizations of the isotropic, unbiased and biased models. We compare in table~\ref{table:comp} the probability distributions of $X_{C,UB}$, $X_{C,B}$, $X_Y(\{\ell\})$, and $X_W(\{D_i\})$ for several sets of $\{\ell\}$ and $\{D_i\}$ where the signal is strongest ($l\le 4$ and $D<60^\circ$). The table presents the probabilities $P(M_1|M_2)$ to rule out model $M_1$ with confidence $y$, assuming the UHECR source distribution follows model $M_2$: $P(M_1|M_2)$ is the probability to obtain, assuming model $M_2$, a value of $X$ ($X_C,X_Y,X_W$) that is obtained in model $M_1$ with probability smaller than $1-y$. Results are presented in the table for the detection of 100 cosmic-rays above $4\times10^{19}$~eV. Clearly, the $X_C$ statistic provides the highest probability to detect the expected anisotropy signal.

\begin{table}
  \centering
\begin{tabular}{|l|ccc|ccc|}
  \hline
    & &95\% CL& & & 99\% CL&\\
    & P(I$|$UB) [\%]  & P(I$|$B) [\%]& P(UB$|$B) [\%] & P(I$|$UB) [\%]& P(I$|$B) [\%]& P(UB$|$B) [\%]\\
\hline
  $X_{C,UB}$ Auger &  23 & 79 & 42 & 5 &  43 & 10\\
    $X_{C,B}$&  26 & 73 & 32 & 7 &  40 & 6\\
\hline
$X_{C,UB}$ Full Sky &  22 & 83 & 48 & 6 &  51 & 19\\
$X_{C,B}$ &  26 & 78 & 37 & 7 &  45 & 9\\
\hline
$X_{C,UB}$ TA &  21 & 82 & 48 & 5 &  46 & 11\\
$X_{C,B}$ &  25 & 78 & 37 & 5 &  37 & 10\\
\hline\hline
$X_W(\{D_i=20,60\})$ Auger &  6 &  9 &  7 &  2 &  3 &  1\\
$X_W(\{D_i=10,20\})$  &  6 & 11 &  9 &  2 &  4 &  2\\
$X_W(\{D_i=0,10,20,30,40\})$ &  7 & 12 & 10 &  2 &  4 &  2\\
\hline
$X_W(\{D_i=20,60\})$ Full sky & 6 &  8 &  7 &  1 &  1 &  1\\
$X_W(\{D_i=10,20\})$&  4 &  7 &  8 &  1 &  1 &  1\\
$X_W(\{D_i=0,10,20,30,40\})$ &  5 &  8 &  7 &  1 &  1 &  2\\
\hline\hline
$X_Y(\{\ell\}=\{1\})$  Auger &  5 &  6 &  6 &  1 &  1 &  1\\
$X_Y(\{\ell\}=\{2\})$   &  6 &  8 &  7 &  1 &  2 &  1\\
$X_Y(\{\ell\}=\{1,2,3,4\})$ &  6 &  7 &  6 &  1 &  1 &  1\\
\hline
$X_Y(\{\ell\}=\{1\})$  Full sky &6 & 10 &  8 &  1 &  2 &  2\\
$X_Y(\{\ell\}=\{2\})$&  6 & 10 &  8 &  1 &  2 &  2\\
$X_Y(\{\ell\}=\{1,2,3,4\})$ &  7 & 11 &  7 &  1 &  2 &  2\\
\hline
\end{tabular}
\caption{ Probabilities $P(M_1|M_2)$ to rule out model $M_1$ at a certain confidence level (CL), assuming that the UHECR source distribution follows model $M_2$, for a given statistic ($X_C,X_Y,X_W$). Numbers in each line are calculated from $10000$ Monte Carlo realizations of 100 cosmic-rays above $40$~EeV. Boxes marked "Full sky" refer to a uniform exposure over the sky, and boxes marked "Auger" ("TA") refer
to an angular exposure pattern corresponding to that of the Auger (Telescope array) experiment. For
this comparison we have used $\bar{s}_0=10^{-4}{\rm Mpc}^{-3}$, $\bar s\propto(1+z)^3$, $d\log
n/d\log E=-2$, and $\Delta E/E=0.2$, and included events along the Galactic disk.
}\label{table:comp}
\end{table}

\begin{table}
\centering
\begin{tabular}{|l|ccc|ccc|ccc|c|}
  \hline
    & &95\% CL& &&99\% CL& && 99.9\% CL &&\\
  $P(M_1|M_2)$ [\%]  & P(I$|$UB)   & P(I$|$B) & P(UB$|$B)   & P(I$|$UB)  & P(I$|$B)  & P(UB$|$B)
    & P(I$|$UB)  & P(I$|$B)  & P(UB$|$B)&$P_R$ \\
\hline
$E>20{\rm\  EeV}$ (402 events) &  26  (31) &  67 (76) &  27 (34) &   7 (12) &  34 (51) &   7 (14) &   $<$1 (3) &  1 (25) &   $<$1 (3) & 99 (16) \\
$E>40{\rm\  EeV}$ 100 events  &    22 (26) &  67 (76) &  30 (38) &   5 (10) &  30 (52) &   7 (16) &   $<$1 (2) &  $<$1 (24) &  $<$1 (4) & 65 (2) \\
$E>60{\rm\  EeV}$ (31 events) &    20 (25) &  60 (73) &  26 (36) &   4 (9) &  23 (48) &   6 (15) &   $<$1 (2) &  $<$1 (19) &   $<$1 (4) & 31 (1) \\
$E>80{\rm\  EeV}$ (10 events) &    15 (19) &  38 (50) &  17 (22) &   3 (6) &  11 (25) &   3 (7) &   $<$1 (1) &  $<$1 (8) &  $<$1 (1) &  9 (0.1) \\
\hline
$E>40{\rm\  EeV}$ $\Delta E/E=0.1$ &   22 (27) &  69 (80) &  33 (41) &   5 (10) &  32 (56) &   9 (18) &   $<$1 (2) &  $<$1 (28) &  $<$1 (5) & 65 (2) \\
$E>40{\rm\  EeV}$ $\Delta E/E=0$ &   23 (27) &  70 (79) &  33 (42) &   5 (10) &  33 (56) &   8 (18) &   $<$1 (2) &  $<$1 (28) &  $<$1 (5) & 64 (2) \\
\hline\hline
$E>20{\rm\  EeV}$ (1205 events) &    45 (62) &  94 (99) &  45 (67) &  13 (36) &  70 (96) &  11 (41) &   $<$1 (12) &  3 (82) &   $<$1 (16) &$>$99.9(90) \\
$E>40{\rm\  EeV}$ 300 events &    39 (54) &  94 (99) &  52 (74) &  10 (29) &  70 (96) &  15 (49) &   $<$1 (10) &  3 (84) &   $<$1 (23) &  $>$99.9(37) \\
$E>60{\rm\  EeV}$ (94 events) &    31 (50) &  87 (99) &  42 (72) &   6 (25) &  44 (94) &   9 (45) &   $<$1 (7) &  2 (78) &   $<$1 (17) &  97 (15) \\
$E>80{\rm\  EeV}$ (31 events) &   22 (36) &  63 (89) &  24 (46) &   3 (16) &  14 (71) &   4 (21) &   $<$1 (4) & $<$1 (39) &  $<$1 (6) & 65 (4) \\
\hline
$E>40{\rm\  EeV}$ $\Delta E/E=0.1$  &  40 (55) &  96 (99.5) &  57 (79) &  10 (29) &  74 (97) &  17 (54) &   $<$1 (10) &  5 (87) &   $<$1 (26) &  $>$99.9(35) \\
$E>40{\rm\  EeV}$ $\Delta E/E=0$   &    39 (55) &  96 (99.5) &  59 (80) &  10 (29) &  74 (97) &  19 (55) &   $<$1 (9) &  4 (87) &   $<$1 (26) &  $>$99.9(35) \\
\hline\hline
$E>20{\rm\  EeV}$ (2410 events) &  62 (86) &  99.4 (99.9) & 63 (90) &  27 (66) & 94($>$99.9) & 24 (72) & 3 (33) & 46  (99.4) & 2 (44) &  $>$99.9($>$99.9) \\
$E>40{\rm\  EeV}$ 600 events & 52 (78) & 99($>$99.9) & 67 (94) & 14 (54) & 88($>$99.9) & 21 (81) & $<$1 (26) &  5 (99.6) &  $<$1 (55) &  $>$99.9(88) \\
$E>60{\rm\  EeV}$ (188 events) & 41 (74) & 95($>$99.9) & 52 (93) & 7 (49) &  61 (99.9) & 12 (78) & $<$1 (21) &  2 (99.1) &  $<$1 (48) &  $>$99.9(60) \\
$E>80{\rm\  EeV}$ (62 events)  & 27 (57) & 76 (99.2) & 29 (69) & 3 (30) &  21 (95) & 5 (42) &  $<$1 (9) &  1 (79) &  $<$1 (16) &  97 (24) \\
\hline
$E>40{\rm\  EeV}$ $\Delta E/E=0.1$   & 53 (80) & 99.2($>$99.9) & 73 (96) & 13 (57) & 90($>$99.9) & 23 (86) & $<$1 (27) &  5 (99.7) &  $<$1 (62) &  $>$99.9(87) \\
$E>40{\rm\  EeV}$ $\Delta E/E=0$  & 54 (80) & 99.3($>$99.9) & 74 (97) & 14 (56) & 91($>$99.9) & 24 (87) & $<$1 (27) &  5 (99.8) &  $<$1 (62) &  $>$99.9(87) \\
\hline
\end{tabular}
\caption{The energy threshold and energy resolution dependence of $P(M_1|M_2)$ for the $X_{C,UB}$
statistic. Model parameters used: $\bar s\propto(1+z)^3$, $d\log n/d\log E=-2$, and $\Delta
E/E=0.2$.  Values are given for $\bar{s}_0=10^{-4}{\rm Mpc}^{-3}$ and in parentheses for
$\bar{s}_0=10^{-2}{\rm Mpc}^{-3}$. An angular exposure pattern corresponding to that of the Auger
experiment was assumed, and the Galactic strip $|b|<12^\circ$ was excluded. Numbers in each line
are calculated from $100,000$ Monte Carlo realizations. The top block of lines gives the results for 100
cosmic-rays above $40$~EeV, the middle block for 300 events, and the bottom block for 600
events (numbers in parentheses in the first column give the corresponding numbers of events above different energy thresholds). The last two lines in each part of the table illustrate the effect of statistical energy uncertainty. The last column gives the probability for the detection of "repeaters", sources producing multiple events, for $\bar{s}_0=10^{-4}{\rm Mpc}^{-3}$ and in parentheses for
$\bar{s}_0=10^{-2}{\rm Mpc}^{-3}$.} \label{table:opt}
\end{table}

\begin{figure*}[h]
   \includegraphics[width=0.9\textwidth]{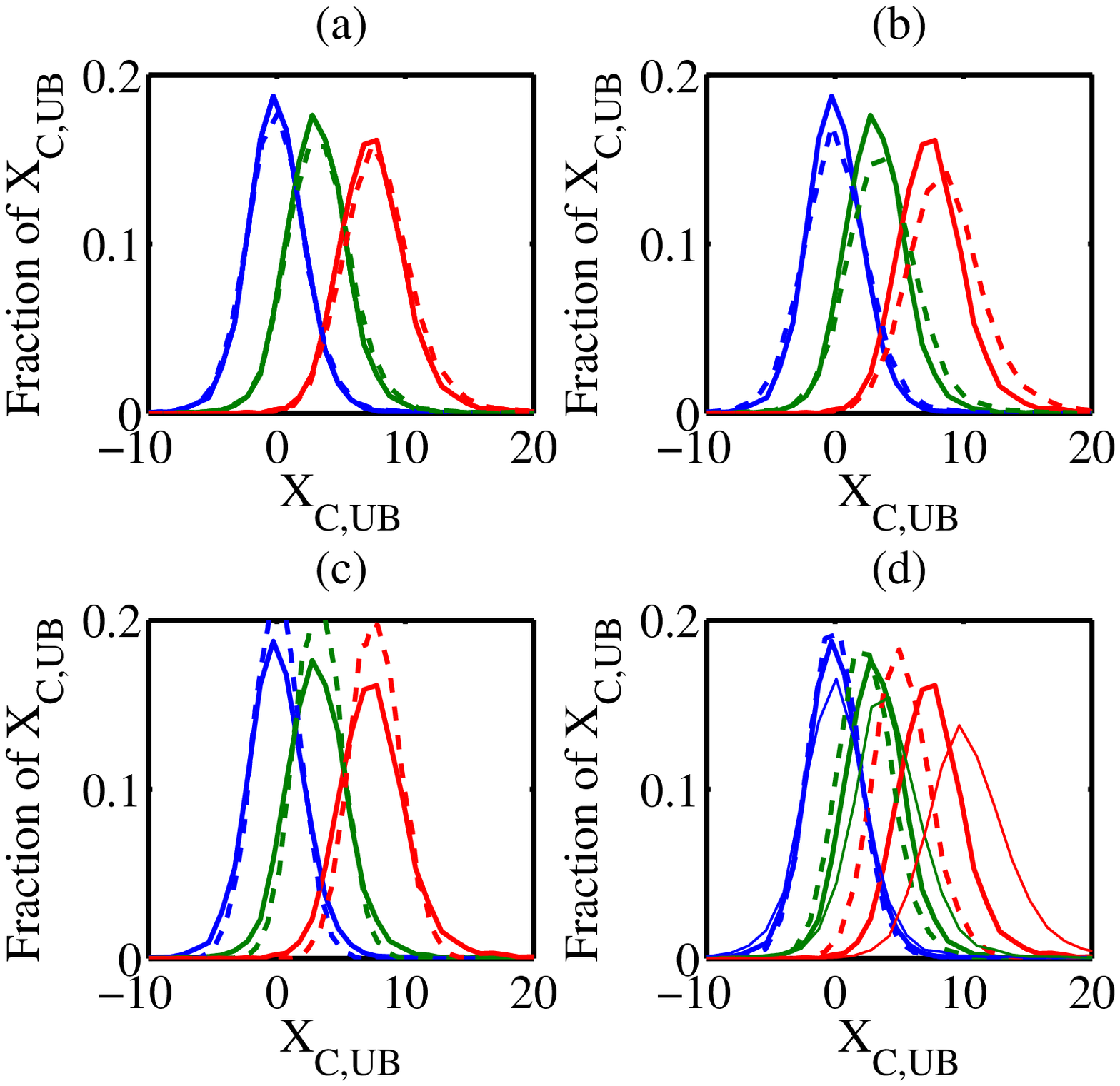}
\caption{The sensitivity of the distribution of $X_{C,UB}$ to variations in model parameters and to systematic
energy uncertainty. All curves obtained from 10000 Monte-Carlo realizations with 300 events above $40$~EeV.
In each panel, solid lines correspond to default model parameters, $\bar{s}_0=10^{-4}{\rm Mpc}^{-3}$,
$\bar s\propto(1+z)^3$, $d\log n/d\log E=-2$, and blue/green/red color corresponds to the isotropic/unbiased/biased model. Panels (a-c) explore the following variations: (a) Dashed lines correspond to $\alpha=d\log n/d\log E=-2.2$; (b) Dashed lines correspond to no source evolution,
$\bar s\propto(1+z)^0$; (c) Dashed lines
correspond to $\bar{s}_0=10^{-2}{\rm Mpc}^{-3}$.
In panel (d) the sensitivity to systematic energy uncertainty is addressed by calculating $X_{C,UB}$
(eq.~\ref{eq:X_C}) with
$N_{i,UB}(E>40{\rm\ EeV})$ and $N_i$ obtained from the simulations for different energy thresholds (and properly normalized): $30$~EeV (dashed line), $40$~EeV (solid line), and $50$~EeV (thin-solid line).}
    \label{fig:sens}
\end{figure*}

Next, we consider the dependence of the strength of the anisotropy signal on the cosmic-ray energy
threshold and on the statistical energy uncertainty. Table~\ref{table:opt} presents the probabilities $P(M_1|M_2)$ for several energy thresholds, $E>20,\ 40,\ 60,\ 80$~EeV, and parameters similar to those chosen for fig.~\ref{fig:comp}, $\bar s\propto(1+z)^3$, $d\log n/d\log E=-2$, $\Delta E/E=0.2$ and angular exposure pattern corresponding to that of the Auger experiment. Probabilities are given for both $\bar{s}_0=10^{-4}{\rm Mpc}^{-3}$ and $\bar{s}_0=10^{-2}{\rm Mpc}^{-3}$. The probabilities were calculated for this table excluding the Galactic strip $|b|<12^\circ$, along which large deflections are expected by the Galactic magnetic field. Comparison of tables~\ref{table:opt} and ~\ref{table:comp} demonstrates that excluding the Galactic strip does not affect the signal significantly. The table demonstrates that the signal is stronger for lower energy. Although the contrast of the fluctuations in the cosmic-ray intensity is higher at high energy (see fig.~\ref{fig:CRmap1}), due to the fact that the propagation distance is smaller, the signal becomes weaker at higher energies since the number of observed UHECR drops rapidly with energy. At $E>20$~eV a significant contamination from Galactic sources may be present and the deflection of cosmic-ray particles may become significant. In order to avoid distortions of the UHECR intensity map by these effects, it is advisable to chose an energy threshold of $E>40$~EeV. The last two lines of each block of table~\ref{table:opt} demonstrate that statistical energy uncertainty of $\Delta E/E<0.2$ does not affect significantly the results.

A comparison of the values of $P(M_1|M_2)$ given in table~\ref{table:opt} for $\bar{s}_0=10^{-4}{\rm Mpc}^{-3}$ and $\bar{s}_0=10^{-2}{\rm Mpc}^{-3}$ shows that the interpretation of the anisotropy signal (i.e. of the measured value of $X_C$) depends on the assumed source number density. While the average value of $X_C$ obtained in realizations of the various models (I, UB, B) is independent of $\bar{s}_0$, its distribution does depend on $\bar{s}_0$: For a lower source number density the number of sources contributing to the flux is lower, hence fluctuations are larger and the distribution of $X_C$ values is wider. This is illustrated in panel c of fig.~\ref{fig:sens}. Thus, in order to determine the probability of obtaining a certain value of $X_C$ in a given model (I, UB, B), one must make an assumption regarding the value of $\bar{s}_0$. As mentioned in \S~\ref{sec:Method}, and discussed in detail in \cite{Waxman:1996hp}, $\bar{s}(z=0)$ may be constrained by the number of "repeaters", the number of sources producing multiple events. The last column of table~\ref{table:opt} presents the probability for observing repeaters at various energies. Comparison of the numbers obtained for $\bar{s}_0=10^{-4}{\rm Mpc}^{-3}$ and $\bar{s}_0=10^{-2}{\rm Mpc}^{-3}$ shows that the presence or absence of repeaters may provide additional constraints on the source density (beyond $\bar{s}_0\gtrsim10^{-5}{\rm Mpc}^{-3}$) once the number of events observed above 40~EeV exceeds several hundreds.

The values of $P(M_1|M_2)$ that appear in table~\ref{table:opt} demonstrate, in accordance with
the results of \cite{Waxman:1996hp}, that the detection of few hundred events above $4\times10^{19}$~eV is required in order to guarantee identification of the predicted anisotropy signature, and discrimination between different bias models, with high statistical significance.

Finally, we consider the sensitivity of the $X_{C,UB}$ distribution to variations in $d\log n/d\log E$, to modifications of the redshift evolution of the source number density, and to uncertainties in the calibration of the absolute energy scale of the experiments. The various panels of  fig.~\ref{fig:sens} compare the distributions of $X_{C,UB}$ obtained for $\bar{s}_0=10^{-4}{\rm Mpc}^{-3}$, $\bar s\propto(1+z)^3$, $d\log n/d\log E=-2$, $\Delta E/E=0.2$ and $E>40$~EeV with those obtained with modified parameters (for all calculations we have used an angular exposure pattern corresponding to that of the Auger experiment and excluded the Galactic strip $|b|<12^\circ$). Panels (a) and (b) demonstrate that the
distribution of $X_{C,UB}$ is not sensitive to variations in  $d\log n/d\log E$ and in the redshift evolution of the source number density. Panel (c) addresses the sensitivity to variations of $\bar{s}_0$, which was already discussed above.

Panel (d) addresses the sensitivity to uncertainties in the calibration of the absolute energy scale. It compares the distributions of $X_{C,UB}$ which are obtained by calculating $X_{C,UB}$ (eq.~\ref{eq:X_C}) with the model average number of events expected for $E>40{\rm\ EeV}$, $N_{i,UB}(E>40{\rm\ EeV})$, and the number of events obtained in the Monte-Carlo simulations for different energy thresholds, $N_i(E>E')$ with $E'=30$~EeV, $40$~EeV, and $50$~EeV (properly normalized). The curve obtained for $E'=30$~EeV ($50$~EeV) simulates the distribution of $X_C$ values that are expected to be obtained for a systematic overestimate (underestimate) of event energies by the experiment. The figure demonstrates
that the distribution of $X_{C,UB}$ values, which is expected to be measured from the distribution of UHECR events with energies inferred {\it by the experiment} to satisfy $E>40$~EeV, is sensitive to
uncertainties of $25\%$ in the experimental calibration of the absolute energy scale: Since the anisotropy is stronger at higher energy, a systematic underestimate (overestimate) of event energies leads to a shift of the $X_C$ distribution towards larger (smaller) values. Since the anisotropy is also stronger (at a given energy) for stronger bias, this implies that an underestimate (overestimate) of the absolute energy scale will lead to an overestimate (underestimate) of the bias of the UHECR source distribution.

\section{Analysis of $>5.7\times10^{19}{\rm\ eV}$ Auger data}
\label{sec:Auger}

\begin{figure*}[tb]
   \includegraphics[width=0.7\textwidth]{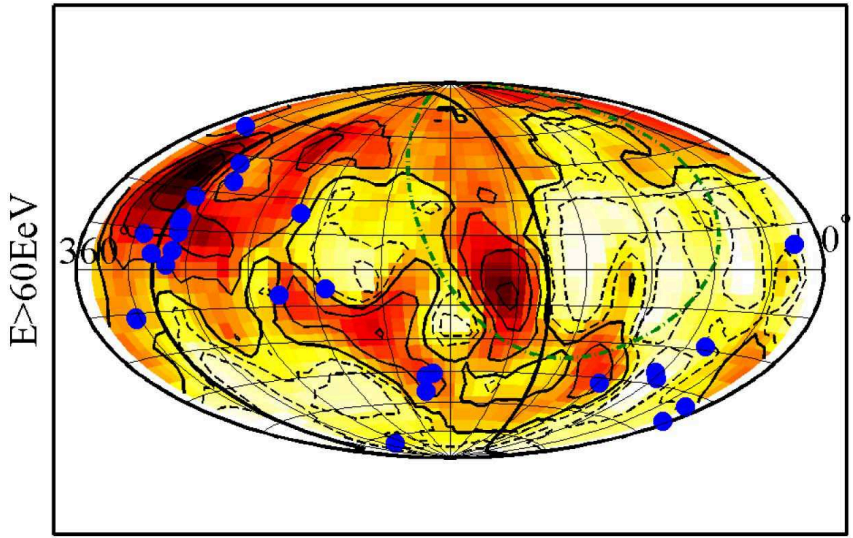}
\caption{The positions of the 27 Auger events with energy exceeding $5.7\times10^{19}$~eV, overlaid on the
intensity map obtained in the biased model (see figure~\ref{fig:CRmap2}).}
    \label{fig:Auger}
\end{figure*}

Figure~\ref{fig:Auger} shows the  positions of the 27 Auger events with energy exceeding $5.7\times10^{19}$~eV, overlaid on the intensity map obtained in the biased model (see figure~\ref{fig:CRmap2}). Examination by eye indicates an enhancement of the UHECR flux associated with the Great Attractor. The significance of this apparent correlation is quantified in table~\ref{table:Auger}. The table presents the value of $X_{C,UB}$ obtained for the Auger data, the average values of $X_{C,UB}$ that are expected to be obtained in the various models (isotropic, unbiased, biased), and the probability that the value of $X_{C,UB}$ obtained for the Auger data would have been obtained in the various models. Since there is a systematic uncertainty in the experimentally determined energy of the events (see fig.~\ref{fig:spectrum}) the table also lists the results obtained under the assumption that the energies of the events are systematically underestimated by 20\%.

\begin{table}
\begin{tabular}{|l|c|ccc|ccc|ccc|}
\hline
 & $X_{C,Auger}$ & $\bar{X}_I$ & $\bar{X}_{UB}$ & $\bar{X}_{B}$& $P_I$&$P_{UB}$&$P_B$\\
\hline
$X_{C,UB}(E>57{\rm\ EeV})$ & 2.7 & 0 & 0.7 & 1.8& 0.9\% (0.2\%)& 3.8\% (2.1\%) & 20.4\% (18.9\%)\\
$X_{C,UB}(E>68{\rm\ EeV})$                  & 3.6& 0& 1.2& 2.9 & 1.5\% (0.1\%) & 5.7\% (2.7\%)& 27.6\% (27.5\%)\\
\hline
\end{tabular}
\caption{ $X_{C,UB}$ obtained for the Auger data, denoted $X_{C,Auger}$, the average values of $X_{C,UB}$
that are expected to be obtained in the various models (isotropic- $\bar{X}_I$, unbiased- $\bar{X}_{UB}$, biased- $\bar{X}_{B}$) for 27 events, and the probability that a value exceeding $X_{C,Auger}$ would have been obtained in the various models (isotropic- $P_I$, unbiased- $P_{UB}$, biased- $P_B$) for 27 events. The 2nd line gives the results under the assumption that the energies of the Auger events are systematically underestimated by 20\% (see fig.~\ref{fig:spectrum}). Model parameters used: $\bar{s}_0=10^{-4}{\rm Mpc}^{-3}$ (and $\bar{s}_0=10^{-2}{\rm Mpc}^{-3}$ in brackets), $\bar s\propto(1+z)^3$, $d\log n/d\log E=-2$, and $\Delta E/E=0.2$. The galactic disk, $|b|<12^\circ$, is excluded from the analysis.}\label{table:Auger}
\end{table}

Based on table~\ref{table:Auger}, the distribution of the arrival directions of $>60$~EeV cosmic-rays reported by the Auger experiment is inconsistent with isotropy at $\simeq98$~\% confidence level, and is consistent with a UHECR source distribution that traces the LSS. According to table~\ref{table:opt}, the probability of ruling out isotropy at $\simeq98\%$ confidence with the observed number of events is not small. The observed distribution indicates a slight preference to a source distribution that is biased with respect to that of IRAS galaxies, that is, there is some indication that UHECR sources are more clustered than IRAS galaxies. As discussed in \S~\ref{sec:Method} and in \S~\ref{sec:results}, the source number density, $\bar{s}(z=0)$, may be constrained by the number of "repeaters", the number of sources producing multiple events. However, although a repeater may be present in the published Auger data, given the possible magnetic field deflections of few degrees (see \S~\ref{sec:B}), the presence or absence of a repeater in the present data does not provide stringent additional constraints on $\bar{s}(z=0)$ (beyond $\bar{s}_0\gtrsim10^{-5}{\rm Mpc}^{-3}$; see table~\ref{table:opt}).

The following cautionary note should be made here. In their search for a correlation between UHECR
arrival directions and the locations of V-C catalog AGNs, the authors of ref.~\cite{AugerSc:2007}
have optimized several search parameters: the maximum angular separation of AGN and UHECR directions, the maximum distance of AGNs included, and the minimum energy of UHECRs included. While the angular separation and AGN distance parameters are irrelevant for our discussion, the minimum UHECR energy might be. As explained in the introduction, it is clear that the anisotropy signature should be searched for in the arrival distribution of UHECRs with energy threshold $\sim5\times10^{19}$~eV. In ref.~\cite{Waxman:1996hp} it was suggested to search for anisotropy with energy thresholds of $4\times10^{19}$~eV and $6\times10^{19}$~eV, and we have made a similar suggestion in sec.~\ref{sec:results} based on table~\ref{table:opt}. The arrival direction data that was made available by the Auger experiment has an energy threshold of $\approx6\times10^{19}$~eV, perfectly consistent with the a-priory energy thresholds suggested. However, the optimization of the energy threshold, to $5.7\times10^{19}$~eV, raises the concern
that this optimization minimizes $P_I$, and hence leads to an overestimate of the significance with which isotropy is ruled out. It is impossible to estimate the magnitude of such an effect based on the data that was made available. In order to confirm the conclusions drawn in the previous paragraph, one may search for the anisotropy signature in the Auger data using the lower energy threshold suggested, $4\times10^{19}$~eV (the probability of detecting a signal at this energy is larger than at higher energy, see table~\ref{table:opt}), or repeat the analysis with an energy threshold of $5.7\times10^{19}$~eV using a larger data set (which will be accumulated on 1~yr time scale).

\section{Summary}
\label{sec:summary}

We have derived the expected angular dependence of the UHECR intensity, under the assumption that the UHECRs are protons produced by extra-Galactic sources that trace the large scale distribution of luminous matter. All sky maps of the UHECR intensity, $I(E,\hat\Omega)$ with several energy thresholds $E$, are presented in figure~\ref{fig:CRmap1} for an unbiased UHECR source distribution, where the UHECR source density is proportional to the LSS density, and in fig.~\ref{fig:CRmap2} for a biased model, where the UHECR source density is proportional to the LSS density in over-dense region and vanishes elsewhere. A numerical representation of the intensity maps may be downloaded from \verb"http://www.weizmann.ac.il/"$\sim$\verb"waxman/criso". The angular structure of the UHECR intensity map reflects the local large scale structure of the galaxy distribution, as can be seen by comparing figures~\ref{fig:CRmap1} and~\ref{fig:CRmap2} to figure~\ref{fig:densityMap75}, which presents the integrated galaxy density out to a distance of 75~Mpc. The anisotropy is larger for higher energy threshold, since the propagation distance increases at lower energy (see fig.~\ref{fig:DiffFlux}) and the contribution to the UHECR flux of sources beyond $\sim100$~Mpc constitutes a roughly isotropic "background". We have used simple analytic arguments (\S~\ref{sec:B}) to show that inter-galactic magnetic fields may modify the intensity map on scales of a few degrees, but not on larger scales (a result which is consistent with detailed semi-analytic and numeric analyses \cite{Dolag:2004kp,Lemoine08}).

We have defined a statistic, $X_C$ (eq.~\ref{eq:X_C}), that measures the correlation between the predicted and observed UHECR arrival direction distributions, and showed that it is more sensitive to the expected anisotropy signature than the power spectrum and the two point correlation function (see table~\ref{table:comp}). The value of $X_C$ for a given data set of UHECR arrival directions, the average  value of $X_C$ over realizations of the various models (isotropic, I, un-biased, UB, biased, B), and the distributions of $X_C$ values expected in realizations of the various models in the limit of infinite UHECR source density, can all be straightforwardly calculated using the numerical representations of the UHECR maps at \verb"http://www.weizmann.ac.il/"$\sim$\verb"waxman/criso". In order to avoid sensitivity to possible distortions of the UHECR intensity map by magnetic fields, we have used $6^\circ\times6^\circ$ bins in calculating $X_C$ and excluded the Galactic plane region, $|b|<12^\circ$. As can be seen by comparing results in tables~\ref{table:comp} and~\ref{table:opt}, the exclusion of the Galactic plane region does not affect the results significantly. Table~\ref{table:opt} demonstrates that the anisotropy signal is stronger at lower energy: Although the contrast of the fluctuations in the UHECR intensity is higher at high energy (see fig.~\ref{fig:CRmap1}), the signal becomes weaker at higher energies since the number of observed UHECR drops rapidly with energy. At $E\sim20$~EeV a significant contamination from Galactic sources may be present and the deflection of cosmic-ray particles may become significant. In order to avoid distortions of the UHECR intensity map by these effects, it is advisable to choose an energy threshold of $E>40$~EeV.

We have shown that the distribution of $X_C$ is not sensitive to the assumed redshift evolution of the source density distribution, to variations in the intrinsic spectrum of protons produced by the sources, and to statistical errors in the experimental determination of event energies (see fig.~\ref{fig:sens} and table~\ref{table:opt}). The distribution of $X_C$ does depend on the assumed local ($z=0$) source number density, $\bar{s}_0$: The average value of $X_C$ obtained in realizations of the various models (I, UB, B) is independent of $\bar{s}_0$, but the width of the distribution is wider, and hence discrimination between models is more diffcult, for lower $\bar{s}_0$ (see fig.~\ref{fig:sens}c and table~\ref{table:opt}). Thus, in order to determine the probability of obtaining a certain value of $X_C$ in a given model (I, UB, B), one must make an assumption regarding the value of $\bar{s}_0$. As mentioned in \S~\ref{sec:Method}, and discussed in detail in \cite{Waxman:1996hp}, $\bar{s}(z=0)$ may be constrained by the number of "repeaters", the number of sources producing multiple events. The last column of table~\ref{table:opt} demonstrates that the presence or absence of repeaters may provide additional constraints on the source density (beyond $\bar{s}_0\gtrsim10^{-5}{\rm Mpc}^{-3}$) once the number of events observed above 40~EeV exceeds several hundreds.

Figure~\ref{fig:sens}d demonstrates that the distribution of $X_{C}$ values, which is expected to be measured from the distribution of UHECR events with energies inferred {\it by the experiment} to exceed a certain threshold, is sensitive to uncertainties in the experimental calibration of the absolute energy scale: Since the anisotropy is stronger at higher energy, a systematic underestimate (overestimate) of event energies leads to a shift of the $X_C$ distribution towards larger (smaller) values. Since the anisotropy is also stronger (at a given energy) for stronger bias, this implies that an underestimate (overestimate) of the absolute energy scale will lead to an overestimate (underestimate) of the bias of the UHECR source distribution. In order to distinguish between different bias models, the systematic uncertainty in the absolute energy calibration of the experiments should be reduced to well below the current $\simeq25\%$.

We have shown, using the $X_C$ statistic, that the recently published $>5.7\times10^{19}$~eV Auger data are consistent with a source distribution that traces LSS, with some preference to an UHECR source distribution that is biased with respect to the galaxy distribution, and inconsistent with isotropy at $\simeq98\%$ CL (see table~\ref{table:Auger}). We have noted, however, that the optimization of the energy threshold made by the Auger collaboration analysis raises the concern that the significance with which isotropy is ruled out may be overestimated (see the end of \S~\ref{sec:Auger}). In order to confirm our detection of a correlation with LSS, one may search for the anisotropy signature in the Auger data using a lower energy threshold, $4\times10^{19}$~eV, or repeat the analysis with an energy threshold of $5.7\times10^{19}$~eV using a larger data set (which will be accumulated on a 1~yr time scale).

According to table~\ref{table:opt}, the detection of $\approx300$ events above $4\times10^{19}$~eV is required in order to enable one to identify the predicted anisotropy signature, and to discriminate between different bias models, with statistical significance exceeding 99\% CL. The experimental exposure required to accumulate this number of events is uncertain, due to the uncertainty in the absolute energy calibration of the UHECR experiments, which implies an uncertainty in the absolute normalization of the UHECR flux (\cite{Bahcall:2002wi}, see also fig.~\ref{fig:spectrum}). Adopting, for example, the absolute energy scale of the HiRes experiment, the average number of events detected above 40~EeV is 300 per year for the Auger exposure of $7000$~km$^2$~sr, and about 50 events per year for the planned Telescope Array exposure of $1200$~km$^2$~sr \cite{TA:2009}. These numbers are reduced by a factor of approximately 2 if the preliminary absolute energy scale of the Auger experiment is chosen (see fig.~\ref{fig:spectrum}). The numbers given in table~\ref{table:opt} imply also that the exposure accumulated within a few years of observation is unlikely to increase the significance of the detection to $>99.9\%$ CL, unless the UHECR source density is comparable (or larger) than that of galaxies.

\acknowledgments This research has been partially supported by ISF, AEC and Minerva grants.

\end{document}